\documentclass[preprint,pra,nofootinbib,superscriptaddress]{revtex4}
\usepackage[mathscr]{euscript}
\usepackage{graphicx}
\usepackage{float,epsfig}
\usepackage{dcolumn}
\usepackage{bm}
\usepackage{graphicx}
\usepackage{amsmath,amssymb,amsthm}
\usepackage{subfigure}
\usepackage{color}
\usepackage[colorlinks=true,linkcolor=blue]{hyperref}
\usepackage[normalem]{ulem}
\usepackage[makeroom]{cancel}

\newcommand{\bea}{\begin{eqnarray}}
\newcommand{\eea}{\end{eqnarray}}
\newcommand{\beq}{\begin{equation}}
\newcommand{\eeq}{\end{equation}}

\def\/{\over}
\definecolor{purple}{rgb}{0.8,0,1}

\begin{document}

\title{Observation of Thouless pumping of light in quasiperiodic photonic crystals}

\author{Kai Yang}
\affiliation{State Key Laboratory of Advanced Optical Communication Systems and Networks, School of Physics and Astronomy, Shanghai Jiao Tong University, Shanghai 200240, China}

\author{Qidong Fu}
\affiliation{State Key Laboratory of Advanced Optical Communication Systems and Networks, School of Physics and Astronomy, Shanghai Jiao Tong University, Shanghai 200240, China}

\author{Henrique C. Prates}
 \affiliation{
 	Departamento de F\'isica and Centro de F\'isica Te\'orica e Computacional, Faculdade de Ci\^encias, Universidade de Lisboa, Campo Grande, Edif\'icio C8, Lisboa 1749-016, Portugal 
}

\author{Peng Wang}
\affiliation{State Key Laboratory of Advanced Optical Communication Systems and Networks, School of Physics and Astronomy, Shanghai Jiao Tong University, Shanghai 200240, China}

\author{Yaroslav V. Kartashov}
 \affiliation{Institute of Spectroscopy, Russian Academy of Sciences, Troitsk, Moscow, 108840, Russia}

\author{Vladimir V. Konotop}
 \affiliation{
 	Departamento de F\'isica and Centro de F\'isica Te\'orica e Computacional, Faculdade de Ci\^encias, Universidade de Lisboa, Campo Grande, Edif\'icio C8, Lisboa 1749-016, Portugal 
}
\author{Fangwei Ye} 
\affiliation{State Key Laboratory of Advanced Optical Communication Systems and Networks, School of Physics and Astronomy, Shanghai Jiao Tong University, Shanghai 200240, China}

\date{\today}

%

\maketitle

{\bf Topological transport is determined by global properties of physical media where it occurs and is characterized by quantized amounts of adiabatically transported quantities.  Discovered for periodic potentials it was also explored in disordered and discrete quasi-periodic systems.  Here we report on experimental observation of pumping of a light beam in a genuinely continuous incommensurate photorefractive quasi-crystal emulated by its periodic approximants. We observe a universal character of the transport which is determined by the ratio between periods of the constitutive sublattices, by the sliding angle between them, and by  Chern numbers of the excited bands (in the time-coordinate space) of the approximant, for which pumping is adiabatic.  This reveals that the properties of quasi-periodic systems determining the topological transport are tightly related to those of their periodic approximants and can be observed and studied in a large variety of physical systems. Our results suggest that the links between quasi-periodic systems and their periodic approximants go beyond the pure mathematical relations: they manifest themselves in physical phenomena which can be explored experimentally.

}

Thouless pumping~\cite{thouless1983quantization}  is an adiabatic process. It is determined by global properties of the physical medium where it occurs, like for example refractive index landscape in the case of wave-guidance, and is characterized by quantized amounts of the transported quantities like charge, matter, or light energy. The ubiquity of transport whose properties are determined by topological properties of the medium, stimulates a permanently growing interest in this phenomenon~\cite{citro2023thouless} across various physical disciplines including  optics~\cite{kraus2012topological, verbin2015topological, zilberberg2018photonic, cerjan2020thouless, wang2022two}, atomic physics~\cite{lohse2016thouless, lohse2018exploring, nakajima2021competition,fu2022nonlinear}, and acoustics~\cite{wauters2019localization}, in addition to the more traditional condensed matter applications~\cite{switkes1999adiabatic, watson2003experimental}.   Discovered for charge currents in periodic potentials~\cite{thouless1983quantization}, the ideas of quantized transport were applied to disordered~\cite{niu1984quantised,wauters2019localization,cerjan2020thouless} and to discrete quasi-periodic~\cite{kraus2012topological,verbin2015topological,cheng2020experimental,nakajima2021competition} physical systems. Already the seminal work of Thouless~\cite{thouless1983quantization}, where the electronic current was considered in a potential consisting of two periodic sublattices moving relative to each other, indicates that the theory should allow for extension to bipartite potentials with incommensurate periods. To date, significant attention was paid to adiabatic pumping in quasicrystals assembled of optical waveguides and described in terms of discrete tight-binding models suitable for deep lattice potentials. The transport in such discrete systems was observed between the edge states~\cite{kraus2012topological,verbin2015topological}. The topology was invoked by mapping to higher (synthetic) dimensional spaces, later being implemented in experiments with 2D photonic waveguides~\cite{zilberberg2018photonic} and cold atoms~\cite{lohse2018exploring}, and allowing for exploration of the physics of the 4D Hall effect.

Previous experimental studies of pumping in quasi-periodic systems were focused on systems allowing for relatively accurate descriptions within the framework of discrete models. The examples are optical lattices for cold atoms, one of which is much deeper than another one~\cite{lohse2016thouless}, or arrays of optical waveguides~\cite{kraus2012topological}. Description of the quantized transport in genuinely continuous quasi-periodic media, which cannot be described within the framework of tight-binding limit (i.e., by a discrete Schr\"odnger equation), but are governed by a continuous Sch\"odinger equation,  like for example relatively shallow lattices, represents several challenges for the theory and experiment. Theoretically, for the phenomenon governed by continuous "time"-dependent ($z$-dependent in optical applications)  Schr\"{o}dinger equation, an  efficient description appealing to higher-dimensional spaces is an open question; the concept of adiabaticity, familiar from the Quantum Mechanics~\cite{messiah2014quantum}, is not applicable straightforwardly (as discussed below); lack of periodicity, infinite spatial extension, and specificity of the spectrum of the continuous models do not allow for introducing conventional Chern numbers~\cite{nakajima2021competition} or Bott indices~\cite{yoshii2021topological} in the time-coordinate space. Experimentally,  the formal definition of a quasi-periodic potential can be given only in the whole space while any optical experimental setting must rely on the modeling using accessible parameters of a finite dimensional system, even when spectrum is addressed~\cite{tanese2014fractal,goblot2020emergence}.

\begin{figure*}[t]
\centering
\includegraphics[width=\linewidth]{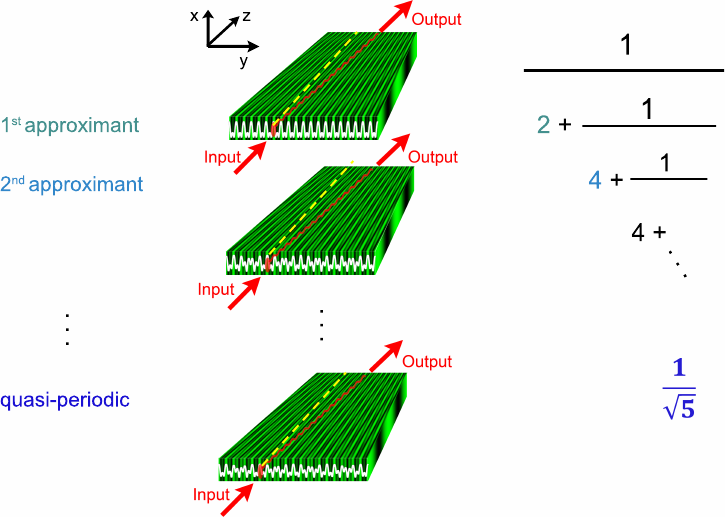}
\caption{Schematics of light pumping in a quasi-periodic photorefractive medium simulated by progressive approximations with periodic lattices.The quasi-periodic crystal is obtained by the superposition of two lattices sliding with respect to each, thus mimicking motion with respect to each other along $y$ direction and homogeneous along $x$ direction (the bottom image in the central column). Such photorefarctive crystal can be treated as a limit of periodic approximants (numbered in the left column and illustrated in the upper images of the central panel) obtained as convergents of the continued fraction representing the BRAs of the irrational relation between the sub-lattice periods (illustrated for $\varphi=1/\sqrt{5}$ in the right column). Considering the beam propagation in these successive approximants and measuring the angles between the input and output beams (i.e., between yellow and red lines in the middle column),  we have found that in a certain approximant the angle reaches a value which remains practically unchanged  in the higher approximations.}
\label{fig:1_s}
\end{figure*}

To overcome these issues, here we elaborate a novel method to tackle the challenges of the study of quantized transport in genuinely continuous quasi-periodic media. By utilizing the method of periodic approximants~\cite{diener2001transition, modugno2009exponential,Marra2020, prates2022bose,zezyulin2022localization,bohr1925theorie}, we investigate the propagation of light in photorefractive crystals created by sliding sublattices, mimicking relative motion at a fixed velocity. Our key insight lies in the use of the best rational approximations (BRAs) ~\cite{khinchin1997continued} to faithfully replicate authentic quasi-periodicity, as illustrated in Fig.~\ref{fig:1_s}. We unveil that starting from a certain order of BRA, subsequent approximations exhibit negligible deviations in dynamics.  The Chern number of a populated band in that approximation, where the conditions of the adiabaticity are still verified, determines the quantized shift of the beam at the output of the quasi-periodic medium.  This allows us to introduce the concept of quasi-adiabatic pumping and quantify transport in quasi-periodic potentials. The displacement of the light beam is shown to be intimately related to the rules of splitting of the band structure into minibands upon increasing BRAs.

Propagation of a paraxial light beam with the amplitude $\Psi$ along the $z$-axis in a photorefractive medium that is homogeneous along the $x$-axis (see schematics in Fig.~\ref{fig:1_s}) is described by the dimensionless Schr\"{o}dinger equation~\cite{efremidis2002discrete,fleischer2003observation}
\begin{subequations}
    \begin{align}
 	\label{main}
 	i\partial_z \Psi=H_\varphi (y,z)\Psi, \qquad \\
 	H_\varphi (y,z)= -(1/2)\partial_y^2-{V_0}/{[1+I_\varphi (y,\varphi y-vz)].}
    \end{align}
\end{subequations}
Here the transverse, $y$, and longitudinal, $z$, coordinates are scaled to the actual beam width $y_0$ (it is $28~\mu \text{m}$ in our experiments) and to the characteristic length $2\pi n_e \lambda$, respectively; $n_e\approx 2.2817$ is the refractive index of the homogeneous SBN: 61 crystal for the extraordinarily polarised light beam at the wavelength of $\lambda=632.8$ nm. The optical potential $V_0/[1+I_\varphi (y,\eta)]$, where $I_\varphi (y,\eta)=p^2|e^{-iy} \cos y+ e^{-i\eta}\cos \eta|^2$ and $\eta=\varphi y-vz$,  is created by the interference of static and moving (sliding) sublattices. The velocity $|v|\ll 1$ is the adiabaticity parameter determined by the tilt angles of the laser beams (in our experiment the $2$ cm sample length accommodates a few pumping cycles).  The $V_0$ stands for the dimensionless d.c. field applied to the crystal (we will use $V_0=-10$ that corresponds to the electric field about $10^5~\text{V/m}$). The irrational number $\varphi$ characterizes quasi-periodicity. The dimensionless lattice amplitude used below is $p=0.5$, which corresponds to an average intensity of the lattice-creating beam $I_\text{av}\approx 1.9~ \textrm{mW}/\textrm{cm}^2$. For more details on the experimental setting, see Materials and Methods. The Hamiltonian $H_\varphi$ is periodic along the propagation direction $z$ with the period $Z=\pi/v$. 

Within the framework of the standard concept of adiabaticity~\cite{messiah2014quantum}, there exist no velocity $v$ for which the transport in a truly quasi-periodic medium can be regarded as adiabatic. Indeed, a typical spectrum of the Hamiltonians $H_\varphi$ with irrational $\varphi$ features a dense fractal-like structure \cite{simon1982almost,yao2019critical} where one can find infinitely close eigenvalues. Among such eigenvalues inter-band energy exchange is unavoidable even for infinitesimal velocities $v$. However, in our model, this coupling is not related to the level crossing, which would induce a finite-dimensional gauge structure~\cite{wilczek1984appearance} and consequently, non-Abelian transport recently observed in~\cite{sun2022non,you2022observation}.  

To characterize pumping in a quasi-periodic medium, we recall that any irrational number $\varphi$ can be expressed as a continued fraction, whose $n$th ($n=0,1,...$) convergent $p_n/q_n$, with $p_n$ and $q_n$ being natural numbers, represents the $n$th order BRA (see e.g.~\cite{khinchin1997continued} and examples in Materials and Methods ). Therefore, we introduce Hamiltonians $H_n$ obtained from $H_\varphi$ by replacing $\varphi$ with $p_n/q_n$ (respectively, $I_\varphi$ is replaced by $I_n$) which are $L_n=\pi q_n$--periodic in the transverse direction. By choosing $n$ large enough one can make  $|I_\varphi-I_n|$  as small as necessary and $L_n\gg\ell$  for any chosen spatial interval $y\in[-\ell/2,\ell/2]$ ( see Materials and Methods).
Consider now propagation of a beam through the $n$-th approximant assuming that the beam width, $\ell_n(z)$, at the input is independent of the BRA order: $\ell_n(0)=\ell_0$. Such beam is governed by $H_n$. If for all $z<Z$ one observes that the beam width and the shift of the center of mass (COM) is small compared to $\ell$, such beam will not "feel" the difference between $I_n$ and its quasi-periodic limit $I_\varphi$. Thus, the idea of our approach is to study the evolution of the beam governed by the subsequent approximants $H_n$, with expectation that the displacement of the COM of the beam will converge to certain limiting value for sufficiently large $n$. Since, each lattice $I_n$ describing $n$-th order approximant is exactly periodic in the $(y,z)$ space, the topological properties of the evolution can be characterized by the Chern numbers calculated for such lattices and defined on the torus $[0,L_n)\times [0,Z)$ associated with each approximant. This idea is illustrated in Fig.~\ref{fig:1_s}.

\begin{figure*}[t]
\centering
\includegraphics[width=11.4cm,height=9.5cm]{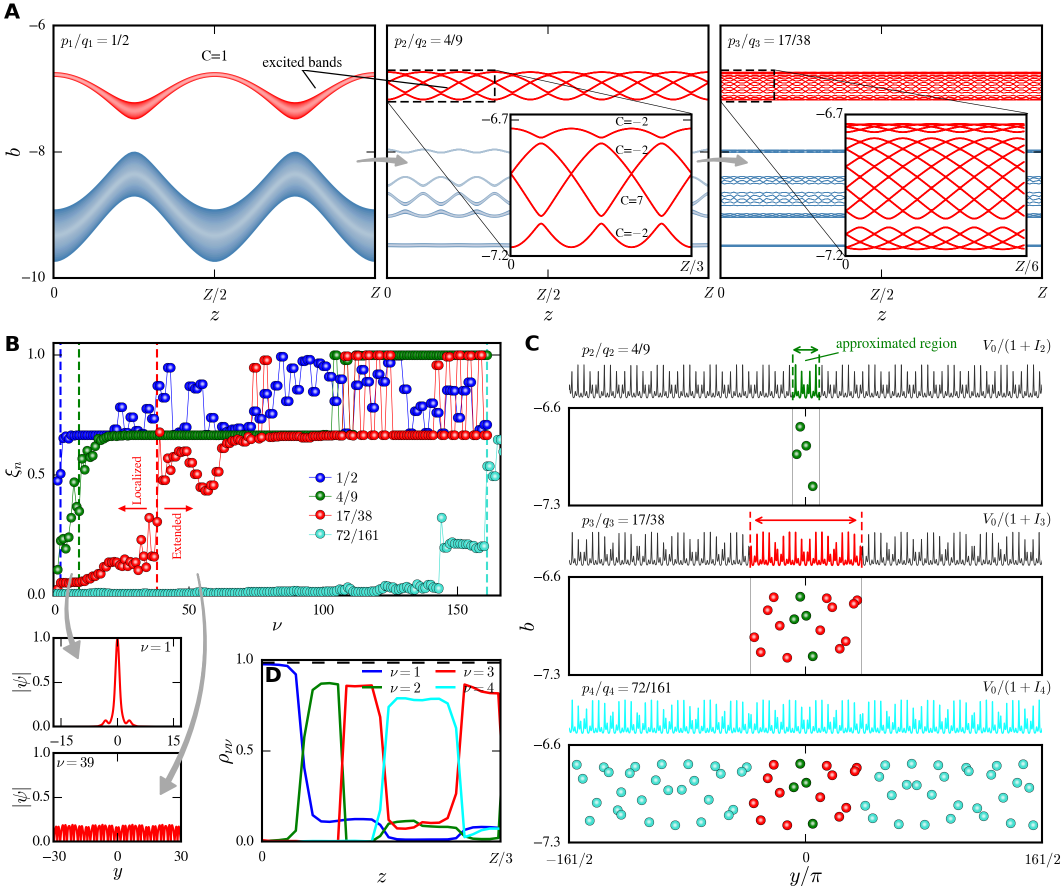}
\caption{Band-gap structure and modes for the lowest approximants of the lattice with $\varphi=1/\sqrt{5}$. (A) Splitting of the bands (from the left to the right)
and the dependence of band edges on propagation distance $z$ for the three lowest BRAs. BRAs and the Chern indexes of the mini-bands that emerged from the upper band of the BRA $p_1/q_1=1/2$ in the leftmost panel (highlighted by red color) are indicated in the panels. The insets show the zooms of the bands. (B) Normalized localization lengths $\xi_n$ of the highest 150 modes for the BRAs indicated in the panel. Different colors distinguish different BRAs, while thin lines are guides for the eye. (C) Distribution of the COMs of the modes (filled circles) belonging to the minibands emerged from the upper band of the first approximation (the red band in leftmost panel (A)), shown in the plane $(y,b)$ for the second, third, and forth  BRAs indicated in the panels. Different colors distinguish the modes inherited from the preceding BRAs and newly appeared in the next one. Thin vertical lines indicate limits of one lattice period $\pm L_n/2$ for each BRA. Each circle corresponds to a different mini-band, i.e., the number of circles in each sub-panel is equal to $p_n$. In panels (B) and (C) are obtained for $z=0.1 Z$. (D) Energy transfer between the two upper bands over one pumping cycle computed for $p_2/q_2=4/9$ approximant. The dashed black line is the total population of the four highest bands. Here and below $V_0=-10$ and  $p=0.5$.}
\label{fig:1}
\end{figure*}

Thus, we consider Thouless pumping in $n$th approximant governed by the periodic Hamiltonian $H_n$ and follow the coordinate of the COM defined by $Y_n(z)=\langle\Psi_n|y\Psi_n\rangle_n/\langle\Psi_n|\Psi_n\rangle_n$
(hereafter $\langle f|g\rangle_n=\int_{\mathcal{I}_n}f^*gdy$ and the asterisk stands for complex conjugation). We concentrate on a beam whose COM 
remains inside the interval $\mathcal{I}_n=[-L_n/2,L_n/2]$, or more specifically: $|Y_n(z)| \ll L_n$. To describe evolution of such beam one can consider an infinite periodic medium. Respectively, we address the "instantaneous" eigenvalue problem for the $n$th periodic approximant: $H_n\phi_{\nu k}^{n}=-b_{\nu k}^{n}(z)\phi_{\nu k}^{n}$ on the whole real axis $y\in\mathbb{R}$.
Here $\phi_{\nu k}^{n}=e^{iky}u_{\nu k}^{n}(y,z)$ is the Bloch function of the band $\nu$, $k\in\mathcal{K}_n=[-1/q_n, 1/q_n)$,  $u_{\nu k}^{n}(y,z)=u_{\nu k}^{n}(y+L_n,z)$, and $b_{\nu k}^{n}(z)$ is the spectral parameter.
Figure~\ref{fig:1}A illustrates how the lowest BRAs of $\varphi=1/\sqrt{5}$ (see Supporting Information) lead to splitting of the upper bands $b_{\nu k}^{n}(z)$ into progressively increasing number of mini-bands.
The highest band in the $(n-1)$th order BRA splits into $p_n$ mini-bands in the $n$th order, the spectrum obtained in the $n$-th BRA holds the memory of spectra of all lower approximants. This phenomenon is one of the manifestations of the "memory effect", whose other properties are discussed below. Thus, in the transition from the $n$-th to the $(n+1)$-th BRA, the number of bands and their individual Chern indices, generally speaking, change. The latter depends on the specific choice of the potential: cf. Fig.~\ref{fig:1} A and Fig. S5 A in the Supporting Information. The mini-bands do not cross although the interval between neighboring bands become as small as $10^{-3}$ already for the $2$nd BRA (making them look like crossings on the scale of the figure). Figure~\ref{fig:1}A reveals also the periodicity of the spectrum of $H_n$ in $z$, which obeys the universal law: $b_{\nu k}^{n}(z)=b_{\nu k}^{n}(z+Z/q_n)$.

Generally speaking the requirement of the relative smallness of the beam width at the crystal output $Z_{\rm out}$: $\ell_n(Z_{\rm out})\ll L_n$, is neither obvious nor valid for the adiabatic evolution if $Z_{\rm out}$ is a sufficiently long distance. Beam diffraction in a periodic medium breaks this requirement. A completely different situation is observed for quasi-periodic potentials whose spectra possess localized modes~\cite{diener2001transition, boers2007mobility, modugno2009exponential, biddle2010predicted, luschen2018single, prates2022bose}. Namely, all eigenstates $\psi_b$ of the problem $H_\varphi\psi_b=-b\psi_b$ are localized in space if $b>b_\textrm{ME}$, where $b_\textrm{ME}$ is the mobility edge (ME)~\cite{mott1987mobility}. These are nondiffracting guided modes (illustrated in Fig.~S3 in Supporting Information). A beam consisting of such modes remains localized along the whole propagation distance. In terms of the periodic approximants, this property manifests itself in localization of the function $u_{\nu k}^n(y,z)$ within each of the intervals $y\in\mathcal{I}_n$ 
for all $ b_{\nu k}^{n}(z)>b_\textrm{ME}$. This is illustrated in Fig.~\ref{fig:1}B where we show the localization length normalized to the lattice period: $\xi_n(\nu,z)=1/(\chi_n L_n)$, for the lowest BRAs  of $\varphi=1/\sqrt{5}$. Here $\chi_n(\nu, z)=\langle (u_{\nu 0}^n)^2|(u_{\nu 0}^n)^2\rangle_n$ 
is the form-factor of the respective mode and Bloch states are considered  normalized.
For computing the form-factor we use the Bloch functions at $k=0$ because with the increase of the BRA order the first Brillouin zone $\mathcal{K}_n$ shrinks, the bands become extremely flat (see Supporting Information), and $u_{\nu k}^n$ becomes weakly dependent on the Bloch wavenumber. In Fig.~\ref{fig:1}B (the modes are numbered from the top to bottom) we observe, that 37 higher modes are well localized already for the $3$rd  approximation, while in the fourth approximation, the number of localized modes above the ME is 161 with 144 higher modes showing particularly strong localization. The  functions $u_{\nu 0}(y,z)$, which are well localized in the interval $y\in\mathcal{I}_n$, accurately approximate the modes in the truly quasi-periodic lattice (see Supporting Information).

Since the number of modes above the ME grows with the order or BRA, tending to infinity in the truly quasi-periodic limit, one  can distinguish two non-commuting limits: (i) $n\to \infty$ at fixed $|v|\ll 1$, and (ii) $v\to 0$ at fixed $n\gg 1$. The latter limit corresponds to standard adiabatic pumping in a one-dimensional system without band crossing. This regime, however, can hardly be implemented for a periodic approximant because the minimal gap width between some mini-bands in the spectrum decreases very rapidly with $n$ (Fig.~\ref{fig:1}A). We have verified numerically that in order to suppress transitions between mini-bands in the relatively low $3$rd BRA the velocity $v$ should be below $10^{-5}$, which would require the usage of unrealistic sample lengths. Thus, we focus on the former limit and define the concept of quasi-adiabatic pumping in a periodic approximant. It should be mentioned, that the velocity as a factor affecting quantized transport, represents a major difference in the manifestation of topology in dynamical Thouless pumping and in the 2D quantum Hall effect~\cite{thouless1982quantized} . In particular, breaking of quantized transport~\cite{privitera2018nonadiabatic} as well as fast Thouless pumping~\cite{fedorova2020observation} have already been addressed in literature. In the case of pumping in a quasi-periodic medium, the role of velocity becomes even more important, because it is intrinsically related to the choice of the BRA dictated by the non-commuting limits of quasi-periodicity and adiabaticity.

We say that a quasi-adiabatic condition is satisfied in the $n$th BRA, if upon pumping all inter-band transitions between the highest and lower bands of $H_n$ are inhibited due to the smallness of  $v$, while transitions between mini-bands emerging in $(n+1)$th BRA from the highest band of $n$th approximation does occur. In simple words, for a given $v$, a quasi-periodic pumping is adiabatic for the highest band of the $n$th approximant but the adiabaticity is violated at $(n+1)$th BRA. This definition reflects the intrinsic non-adiabaticity of the pumping in continuous quasi-periodic media.
Thus, we study the quantized transport carried by localized modes for increasing BRA orders at fixed $|v|\ll 1$ in samples having lengths of approximately two pumping cycles $Z_\textrm{out}=2Z$. 

Fig.~\ref{fig:1}C illustrates the localization of the eigenmodes of $H_n$ in the energy (or in propagation constant) and in coordinate spaces calculated using the Bloch modes at $k=0$ and arbitrarily chosen distance $z=0.1Z$. One observes that the eigenstates belonging to the newly created mini-bands in the $(n+1)$th BRA approximation appear inside the interval $\mathcal{I}_{n+1}$ but outside the interval $\mathcal{I}_{n}$ (which is illustrated by colors). At the same time, the modes that emerged in the $n$th approximation remain only weakly affected in the $(n+1)$th approximation in accordance with a negligible variation of the potential (as discussed above). In other words, at each new BRA, the system keeps memory about energy and localization position of the modes obtained in the preceding approximations.  This  memory effect, verified also for other quasi-periodic media~\cite{zezyulin2022localization,prates2022bose}, provides yet another explanation of why after certain order of BRA, the approximations of the higher orders do not affect the evolution of the light beam over a few cycles. Indeed, when $z$ increases the modes adiabatically "move" in the $(y,b)$-space, although never colliding with each other because of avoided crossing. For the energy exchange between localized modes, they must be "close" enough not only in energies (propagation constants) but also in the coordinate space. Hence, a mode excited at the input, $z=0$, far from the boundaries of the interval $\mathcal{I}_n$, requires extremely long propagation distances in order to exchange its energy with modes outside the interval $\mathcal{I}_n$. This means that the modes added in the interval $\mathcal{I}_{n+1}$ upon next $(n+1)$th BRA do not affect the evolution of the majority of the modes from the interval $\mathcal{I}_n$ (except a few modes located close to the boundary of that interval). This explains also relatively rare  instants of the energy exchange, illustrated in Fig.~\ref{fig:1}D where the quasi-adiabatic condition is satisfied for the 1st BRA of $\varphi=1/\sqrt{5}$.

The aforementioned predictions based on the analysis of the spectrum of the periodic approximants, have been tested in two separate experiments reported below in Figs.~\ref{fig:2} and~\ref{fig:3}. In these experiments, we utilize the technique of optical induction to create in a photorefractive crystal SBN:61 with dimensions $5 \times 5 \times 20$ \text{mm}$^3$  Materials and Methods a series of periodic lattices representing approximants of progressively increasing order $n$ for quasi-periodic lattices described by two irrational numbers $\varphi$. To probe the induced lattices we used a low-power extraordinarily polarised laser beam that passed first through a cylindrical lens to form a quasi-one-dimensional beam which is uniform in the $x$ direction and has finite width of $28~\mu \text{m}$ in the $y$-direction. This beam was launched normally onto the front facet of the crystal at the position of one of the local lattice maxima. It covered approximately one lattice stripe clearly visible in Fig.~\ref{fig:2}A. The beam excites at the input of the crystal the highest mini-bands (discussed in the figures). The transformation of the beam within the volume of the sample is captured using a $z$-shifting CCD  Materials and Methods.

First, we consider the pumping of light in approximants of the quasi-periodic lattice with $\varphi=1/\sqrt{5}$. Fig. \ref{fig:2}A shows intensity distributions of the lattice wave field at different propagation distances within one longitudinal period 
obtained at 4th BRA ($\varphi_4=72/161$, which differs from the actual value 
$\varphi=1/\sqrt{5}$ by only $0.003\%$). It is uniform in $x$, modulated in $y$, and replicates itself after each longitudinal period (in the physical units $Z\approx 8~\text{mm}$). In Fig.~\ref{fig:2} B and C we illustrate the pumping dynamics by presenting experimentally measured COM displacement $Y(z)$ of the probe beam in the lattice from Fig.~\ref{fig:2}A and evolution of intensity distribution of the beam in the $(y,z)$ plane. One can see that the beam exhibits moderate diffraction, while its COM shows pronounced displacement in the positive direction of the $y$-axis that substantially exceeds the width of the input beam. Comparison of experimentally measured COM displacement with distance $z$ in lattices corresponding to different BRA orders is presented in Fig.~\ref{fig:2}B. Figure~\ref{fig:2}D shows output COM positions for different approximations at the distance $2Z$. We observe that while the 1st BRA provides only a rough estimate for pumping dynamics in the authentic quasi-periodic lattice, the COM displacement rapidly reaches saturation, as predicted above (visible already in the $3$rd BRA), that allows us to conduct experimental observation of pumping within $2$ cm-long sample.

To characterize the observed pumping quantitatively, i.e., to relate it with the Chern numbers of the periodic approximants, we assume that the quasi-adiabatic approximation is verified for the $m$th BRA approximation and consider an $n$th BRA of a higher order ($n>m$) at which the highest band of the $m$th BRA is split into $N_n^m$ minibands of the $n$th BRA. If the power of the input field $U$ is initially launched (fully contained) in one (or a few) of these mini-bands, then the pumped field can be searched in the form $\Psi=\sqrt{U}\sum_{\nu=1}^{N_n^m}\int_{\mathcal{K}_n}c_{\nu k}^n(z)\phi_{\nu k}^{n}dk$, where $z$-dependent coefficients $c_{\nu k}^n(z)$ determine evolution of the power distribution among the modes. 
 
Let us define two Hermitian $N_n^m\times N_n^m$ matrices: the density matrix  ${\bm \rho}_n$ with the entries $\rho_{\mu\nu}^n(z)=({2\pi}/{L_n})[c_{\nu 0}^n]^*c_{\mu 0}^n$ (see Supporting Information), and the Chern matrix ${\bf C}_n$, with the elements $C_{\mu\nu}^{n}(Z)=\frac{i}{2\pi} \int_0^Z dz \int_{\mathcal{K}_n}dk(\langle {\partial_z u_{\mu k}^{n}}|\partial_k u_{\nu k}^{n}\rangle_n-\langle\partial_k u_{\mu k}^{n}|\partial_z u_{\nu k}^{n}\rangle_n)$. The diagonal elements of ${\bf C}_n$, which we denote as $C_\nu^m=C_{\nu\,\nu}^m$, are the Chern indices of the mini-bands involved in pumping computed in the $(y,z)$ plane. Considering $Y_n(0)=0$ (what corresponds to our experimental data), one can obtain an approximate expression describing COM displacement in the $n$th BRA (see Supporting Information)
 $
   Y_{n}(Z) =L_n \mbox{Tr}\{ {\bm \rho}_n(Z){\bf C}_n(Z) \}.
 $  
By construction $\mbox{Tr}\{{\bm \rho}_n(Z)\}=1$ and 
$\mbox{Tr}\{ {\bf C}_n(Z)\}=C_1^m$, where $C_1^m$ is the Chern number of the band of the $m$th BRA for which quasi-adiabaticity is verified (recall that splitting of the highest band with $\nu=1$ is considered). 

For the examples shown in Figs.~\ref{fig:2} and ~\ref{fig:3}, when the quasi-adiabatic condition is verified at the 1st BRA, i.e. for $m=1$, we obtained $N_n^1=p_n$.  At each order $n$ of BRA only two different values of Chern numbers can appear: the diagonal elements of ${\bf C}_n$ acquire either $(-1)^n q_{n-1}$ or $(-1)^n (q_{n-1}-q_n)$ values. This link of the values of Chern numbers to the "history" of obtaining periodic approximant represents yet another manifestation of the memory effect discussed above (Fig.~\ref{fig:1}C), which is consistent with the fact that BRAs (considered here) are convergents of the infinite continued fraction~\cite{khinchin1997continued} (see also schematics in Fig.~\ref{fig:1_s}). In the example at hand  $\mbox{Tr}\{ {\bf C}_n(Z)\}=C_1^1=1$. This is the Chern number defining one-cycle displacement of the beam in the respective quasi-periodic potential.

While the introduced matrices ${\bf C}_n$ are determined by the topological properties of the periodic approximants, the matrix  ${\bm \rho}_n$ depends on the field of the input beam and on the magnitude of pumping velocity. Since for $(n+1)$th approximant the adiabaticity is broken, the dependence $Y_{n}(Z)$  can be computed only numerically by considering the evolution governed by Eq.~(\ref{main}).  Meantime, exploring similarities between   properties quasi-periodic and disordered media, one may conjecture that in the course of sufficiently long evolution the populations of the mini-bands become nearly equal with relatively weak dispersion. Subject to this conjecture, the density matrix at the output is expected to have one eigenvalue of order one and all other eigenvalues negligible. Respectively,  the displacement $Y_{n}(Z)$ can be approximated as: $Y_n(Z)\approx ({L_n}/{N_n^m})C_1^m$. In the quasi-periodic limit and for a finite, but sufficiently large $Z$, we obtain the limit for the COM displacement after one period as $Y_n(Z) \to Y_\varphi (Z)$, where
\begin{align}
\label{analyt}
    Y_\varphi(Z)=L_\varphi C_1^m , \qquad L_\varphi=\lim_{n\to\infty}\frac{L_n}{N_n^m},
\end{align} 
and $m$ is the order of BRA for which the pumping is quasi-adiabatic. We emphasize that, although calculations are performed for a chosen order of BRA, the result (\ref{analyt}), viewed as a limit $n\to\infty$ corresponds to authentic quasi-periodic limit, which is achieved already at relatively low BRAs. Although these conclusions based on the conjecture, below we will see that it is confirmed in experiment and its numerical modeling. 

Applying Eq.~(\ref{analyt}) to the case reported in Fig.~\ref{fig:2} we obtain $Y_\varphi(Z)=\pi\sqrt{5}$ (now $N_n^m=p_n$). In Fig.~\ref{fig:2}B the respective trajectory is marked by the dot-dashed lines and in panel B the output displacement $Y_n(2Z)$ is shown by black filled boxes. We observe that the conjecture (\ref{analyt}) agrees well with the experimentally observed COM displacement, especially after two pumping cycles when energy redistribution among the bands is more equilibrated. Meantime, the experimental COM displacement is slightly lower than the theoretically predicted value of $Y_{n}(Z)$ ({Fig.~\ref{fig:2}D}). We attribute this to the partial decay of the light beam intensity due to the radiation, i.e., to the excitation of delocalized modes in the experiment, as well as the presence of small and irregular background noise that occurs unavoidably in experiments. This small background noise consists of two components. The first component is caused by the scattering of light from the laser by the atmosphere,  which appears as ambient light. The second component is the weak scattering that occurs when the laser beam enters the crystal which is also randomly distributed  throughout the crystal. Therefore, by disregarding  a certain amount of light side lobes, the agreement between theory and experiment can be enhanced. This is shown in the Fig.~\ref{fig:2}D,  where the curve “experiment 1” stands for the calculation of the COM from the whole light field, while “experiment 2” stands for the calculation of the COM by discarding the contribution of the field below 20\% of the peak intensity (see also Fig.~\ref{fig:3}D).

The second set of experiments was performed with the approximants of the quasi-periodic lattice with $\varphi=(\sqrt{5}+1)/4$. Although with different specific values, the memory effect is fully confirmed in this case too (see Materials and Methods for the BRAs and Fig.~S5 for the characterization of the spectrum, and mode localization).  Now the periods of the sliding sublattice substantially differ from the periods in the experiments of Fig.~\ref{fig:2}. The lattice approximant in the $4$th order BRA, $\varphi_4= 72/89$ (its accuracy of approximation of the irrational number $\varphi=(\sqrt{5}+1)/4$ is $0.002\% $), is presented in Fig.~\ref{fig:3}A, the trajectory of COM of the beam propagating in such lattice is shown in Fig.~\ref{fig:3}B, while propagation dynamics is illustrated in Fig. \ref{fig:3}C. Since now the period of the approximating lattice is nearly two times smaller than the period of lattice in Fig.~\ref{fig:2}, we observe more pronounced diffraction of the beam during the pumping process (cf. Fig.~\ref{fig:3}C and Fig.~\ref{fig:2}C). In this second set of experiments, the Chern number of the upper band of the 2nd BRA, at which the quasi-adiabatic approximation holds (i.e. now $m=2$ according to Fig. S5 in Supporting Information ), is found to be $C_1^{2}=-1$. Now the light is pumped in the negative $y$ direction, which is opposite to the positive direction of motion of the sliding lattice Materials and Methods. This again clearly indicates on the topological nature of pumping.

The band splitting upon the increase of BRA in this latter case occurs according to the rule $N_n^2=q_n-p_n$, that for $L_n=\pi q_n$ yields $L_\varphi=\pi/(1-\varphi)$  in (\ref{analyt}). The respective predictions for COM displacement are shown in Fig.~\ref{fig:3}B by dot-dashed line and in panel Fig.~\ref{fig:3}D by filled boxes.  Thus, we observe that the displacement predicted by the conjecture (\ref{analyt}) is about 15\% larger than the  displacement observed in the experiments. We also observe that the saturation with the increase of the order of BRAs is not satisfactory yet. This is the experimental limitation due to the finite size of the photorefractive crystal: for observing the effects appearing due to the next approximant of 5th BRA $q_5=377$ (Eq.~(1b) in Supporting Information Materials and Methods) one would need a much longer and wider crystal to accommodate two pumping cycles for much smaller velocities and the possibility of the beam go beyond the interval $\mathcal{I}_4$, although remaining inside $\mathcal{I}_5$.  At the same time, one still observes a rapid and stable tendency to the saturation of the displacement with an increase in the order of BRA, thus, corroborating with the above predictions. 

 The described pumping scenario, i.e., the shift of the COM of the beam over one cycle, does not depend on the number of excited miniband, provided it is below the ME. In Fig.~S7 of the Supporting Information, we provided additional results of simulations for the second and eighth miniband excitation. While the higher-miniband modes, as illustrated in Fig.~S2 B, exhibit profiles with two humps, the trajectories of their COMs follow the same direction as those corresponding to the excitation of the first miniband, shown in Fig.~S7 B.

It is also worth noting that the described pumping scenario is not influenced by the specific details of the underlying lattices, as long as their propagation constants remain below the ME. This is demonstrated by simulations showing the transport of a light beam under varying lattice modulation depths, i.e., of $V_0$. As illustrated in Fig.~S8 in the Supporting Information, increasing $V_0$  significantly suppresses diffraction due to the enhanced waveguide confinement effect. Nonetheless, despite these changes, the COM of the beam remains unaffected, and its behavior is accurately predicted by the previously discussed theory.

\subsection*{Discussion}

We reported on the experimental observation of the topological pumping of light in continuous lattices that are periodic approximants of quasi-periodic optical potentials. This unusual system, where even arbitrarily slow pumping remains non-adiabatic in the truly quasi-periodic limit, nevertheless allows for observation of a universal quasi-adiabatic scenario. The beam displacement is determined by the topological properties of the approximants obtained using the best rational approximations for the irrational relation between periods of the constituent sublattices. While the consideration was performed for fixed pumping velocities, one can conjecture what happens if the velocity is reduced significantly while a truly quasi-periodic lattice is considered. Upon such change when higher best rational approximations become quasi-adiabatic the Chern index defining the shift of the beam after one cycle is changed and hence, the output position of the beam center of mass is changed, too. Thus our observations imply that the fundamental mathematical problem of best rational approximations of irrational numbers allows for experimental visualization. On the other hand, quasi-periodic potentials and their rational approximants, created and discussed here in photorefractive crystals, are obtainable in a variety of other physical settings, including systems of cold atoms, polariton condensates, or arrays of optical waveguides, thus indicating the generality of the observed phenomenon.  Furthermore, the current scheme for wavepacket pumping was successfully implemented in a lattice where quasi-periodicity occurs solely in one dimension, therefore an intriguing avenue for further exploration lies in the generalization of this approach to two- and three-dimensional incommensurate systems, such as dynamic moir\'e lattices. In such systems, one can anticipate the fascinating possibility of engineering the precise topological routing of wavepackets within higher-dimensional spaces with aperiodicty.

\section*{Methods}

\subsection*{Experimental setup} 
The experimental setup is depicted in Extended Data Fig. 1. The lattice was created using the optical induction technique, as described in Ref.~\cite{efremidis2002discrete,fleischer2003observation}. A continuous-wave, frequency-doubled Nd: YAG laser operating at a wavelength of $\lambda=532\text{nm}$ and ordinarily polarized after passing through a half-wave plate (HWP),  was employed to "write" superlattices in a biased photorefractive crystal (SBN: 61). The crystal, with dimensions of $5\times 5\times20~\text{mm}^3$, was oriented such that the optical axis was defined by the $20~\text{mm}$ direction (indicated by the dashed line in the beam path $\bm a$). To investigate the light propagation dynamics in the induced superlattice, an extraordinarily polarized beam at a wavelength of $\lambda=632.8\text{nm}$, generated by a He-Ne laser, was directed into the sample (as shown in the beam path $\bm b$). 
A translation stage equipped with an imaging lens ($L_3$) and a mounted CCD camera enables the recording of induced optical lattices and the intensity of the probe beam. The recording was done step by step, every 1~mm, throughout the sample.

\subsection*{Configuration of the writing beams}
To generate a periodic approximate by superimposing a static lattice with a sliding one, the lattice-writing beam, after being expanded by a spatial filter (SF), sent through an amplitude mask. The amplitude mask has three pinholes on a line, as shown in the inset of Extended Data Fig. 1. Pinholes 1 and 3 are symmetrically positioned with respect to the optic axis, while pinhole 2 is located between them but displaced from their midpoint (not on the optic axis). The distance between pinhole 2 and pinhole 1 is $p_n/q_n$ times the distance between pinhole 3 and pinhole 1, 
 $p_n/q_n$ being the $n$th BRA of an irrational number $\varphi$. In the experiment, we explored $\varphi = 1/\sqrt{5}$ and $(\sqrt{5} + 1)/4$.
In the conventional notations where the leftmost number corresponds to the zero-order BRA, while each subsequent number from the left to the right corresponds to the passage from $n$th to $(n+1)$th BRA, the BRAs obtained as convergents of the continued fractions~\cite{khinchin1997continued}, are 
\begin{subequations}
\begin{align}
\label{BRA1}
     \frac{1}{\sqrt{5}}=[0; 2, 4, 4, ...], \quad   \frac{p_n}{q_n}= \left[0, \frac{1}{2},\frac{4}{9},\frac{17}{38}, \frac{72}{161}, \dots\right]
\\
\label{BRA2}
    \frac{1+\sqrt{5}}{4}=
    [0; 1, 4, 4, ...],\quad  \frac{p_n}{q_n}= \left[0, \frac{1}{1},\frac{4}{5},\frac{17}{21}, \frac{72}{89},  \frac{305}{377}, \dots\right].
\end{align}
\end{subequations}
Notice that $p_n>p_{n-1}>\cdots$ and $q_n>q_{n-1}>\cdots$ and $n=0,1,...$.

The two beams originating from pinholes 2 and 3 are attenuated (AT), such that their amplitude becomes two times smaller, and then interfere with the beam originating from pinhole 1 (which is unattenuated and has an electric amplitude $E_0$) after being refracted by the lens $L_1$. The resultant interference intensity $I(\bm{r})$ is given by,
    \begin{equation}
    \label{eq-I}
        I =\frac{E_0^2}{4} \left| 2e^{i \bm{k}_1  \cdot \bm{r} } + e^{i \bm{k}_2 \cdot \bm{r} } + e^{i \bm{k}_3 \cdot \bm{r}}  \right|^2
    \end{equation}
where $\bm{r}=(y, z)$, $\bm{k}_m$ ($m=1,2,3$) is the wavevector of the $m$th plane wave originating from the $m$th pinhole, each having a transverse ($y$) and a longitudinal ($z$) components,  
    \begin{subequations}
    \label{eq-k}
        \begin{align}
         \bm{k}_1 &= \left( k_y, \sqrt{k_0^2 - k_y^2} \right) \\
            \bm{k}_2 &= \left( -k_y, \sqrt{k_0^2 - k_y^2} \right)\\
            \bm{k}_3 &= \left(  \left(1-2\frac{p_n}{q_n}\right) k_y, \sqrt{k_0^2 - \left(1-2\frac{p_n}{q_n}\right)^2 k_y^2}  \right)
        \end{align}
\end{subequations}    
By substituting Eqs.~(\ref{eq-k}) into Eq.~(\ref{eq-I}), one finds that the resultant intensity of the interference is composed of two lattice components, namely $I = |V_{\text{static}}(y) + V_{\text{sliding}}(y, z)|^2$, where:
   \begin{subequations}
        \begin{align}
        \label{stat}
            V_{\text{\text{static}}}& = E_0 e^{-ik_y y}\cos(k_y y), 
            \\
            \label{slide}
            V_{\text{\text{sliding}}}& =E_0 e^{-i(p_n/q_n)k_y \eta}\cos\left(\frac{p_n}{q_n}k_y \eta\right), \quad 
            \eta=  y-\left(1-\frac{p_n}{q_n}\right)\frac{k_y}{k_0} z
        \end{align}
    \end{subequations} 
     
Thus, in our experiment, the spatial period of the static lattice, defined by Eq.~(\ref{stat}) in the physical units  as $\pi/k_y$ is fixed. It is  $15~\mu\text{m}$ in all
experiments emulating quasi-periodic lattice
with $\varphi=1/\sqrt{5}$, and $13~\mu \text{m}$ in those emulating quasi-periodic lattice with $\varphi=(1+\sqrt{5})/4$. The transverse period of the sliding lattice, $\pi q_n/(k_yp_n)$,  as well as its longitudinal period of the superlattice depend on the order of the BRA. Specifically, here we provide the relevant parameters for the first fourth BRAs for $\varphi=1/\sqrt{5}$: our experimentally induced sliding lattices have the transverse periods $30.00~\mu \text{m}, 33.75~\mu \text{m}, 33.53~\mu \text{m}$, and $33.54
~\mu \text{m}$, respectively. The longitudinal period corresponding to these approximations are $7.81~\text{mm}, 8.07~\text{mm}, 8.01~\text{mm}$, and $ 8.01~\text{mm} $, respectively. Already at the 4th BRA, the sliding lattice shows no discernible variation on the transverse scale of the specimen, with BRAs of higher orders. Finally, the transverse period of the superlattice is measured to be $30~\mu\text{m}, 136~\mu\text{m}, 573~\mu\text{m}, 2.40~\text{mm}$, which all agree very well with the theoretically predicted value, $\pi q_n/k_y$.  

\subsection*{Boundary conditions}
In our experiment, the sample has a fixed transversal size of 5 mm × 5 mm throughout all experiments, and the initial beam width is also fixed at 28 µm too. The width of the light beam does change during propagation in the sample due to diffraction effect. However, despite diffraction effect, after propagating over 2 cm within the sample, the beam width only increases to a few hundred micrometers for the light propagation in the fourth rational approximant in our experiment (Fig. S9 A and B). This means that the waist of the light beam remains significantly smaller than the sample width, which ensures the validity of theoretical explanation of the observed results.

\section*{Data availability}
The data that support the findings of this study are available from the corresponding author upon reasonable request.

\bibliography{natureletter-LT}

\paragraph*{\bf Acknowledgments} 

This work was supported by  the Portuguese Foundation for Science and Technology (FCT) under Contracts UIDB/00618/2020 (DOI: 10.54499/UIDB/00618/2020) and PTDC/FIS-OUT/3882/2020 (DOI: 10.54499/PTDC/FIS-OUT/3882/2020), the FCT doctoral grant 2022.11419.BD, Shanghai Outstanding Academic Leaders Plan (No. 20XD1402000), the National Natural Science Foundation of China (No. 12404385), China Postdoctoral Science Foundation (No. BX20230218, No. 2024M751950) and FFUU-2024-0003 of the Institute of Spectroscopy of the Russian Academy of Sciences.

\paragraph*{\bf Author contributions}

Kai Yang contributed equally to this work with Qidong Fu and Henrique C. Prates.All authors contribute significantly to the work.

\paragraph*{\bf Competing interests} The authors declare no competing interests.

\begin{figure*}[h]
\centering
\includegraphics[width=\linewidth]{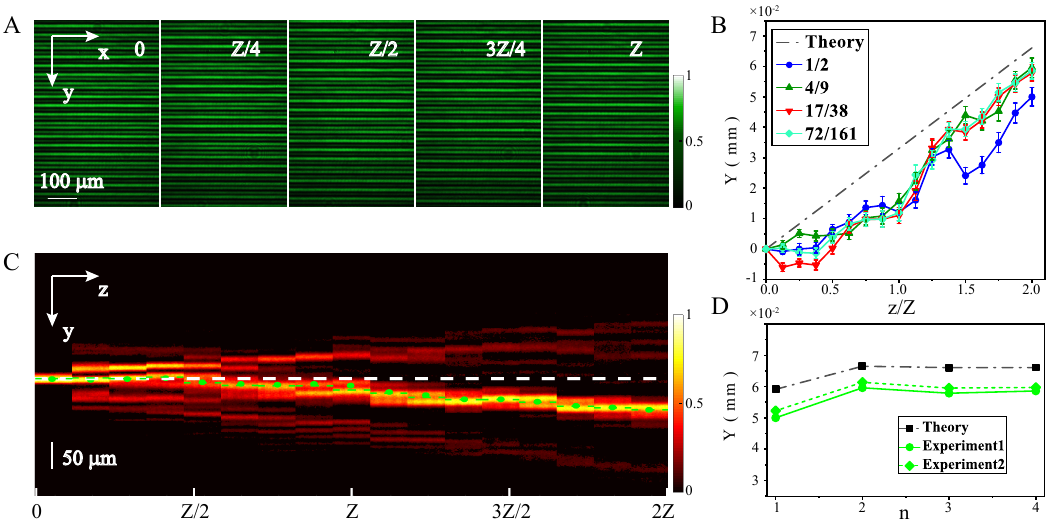}
\caption{ Light pumping in the approximant of a quasi-periodic lattice with $\varphi=1/\sqrt{5}$. (A) The profile of optically induced lattice corresponding to the $4$th order BRA ($p_4/q_4=72/161$) of a quasi-periodic lattice with $\varphi=1/\sqrt{5}$ at different distances $z=0,~Z/4,~Z/2,~3Z/4$ and $Z$ within one pumping cycle. (B) Experimentally measured COM position {\it versus} distance $z$ (dots) compared with the analytical prediction (\ref{analyt}) shown by the dot-dashed line.  (C) The intensity distribution of the probe beam
measured with the step of $Z/8$ ($\approx1$ mm) along the $z$-axis. The green dots superimposed on the plot indicate the positions of the COM for each $Z/8$ interval, while the white dashed line indicates the $y=0$ position, where the incident beam was launched.  (D) The experimentally measured (green) and predicted by (\ref{analyt}) COM coordinate at the output $z=2Z$ for the first four BRAs $n$. Curve “experiment 1” stands for the calculation of the COM from the whole diffraction field, while “experiment 2” stands for the calculation of the COM by discarding the contribution of the field below 20\% of the peak intensity. In all cases, $Z\approx 9.5~\text{mm}$, although it slightly varies with the order of rational approximations.}
\label{fig:2}
\end{figure*}

\begin{figure*}
\centering
\includegraphics[width=\linewidth]{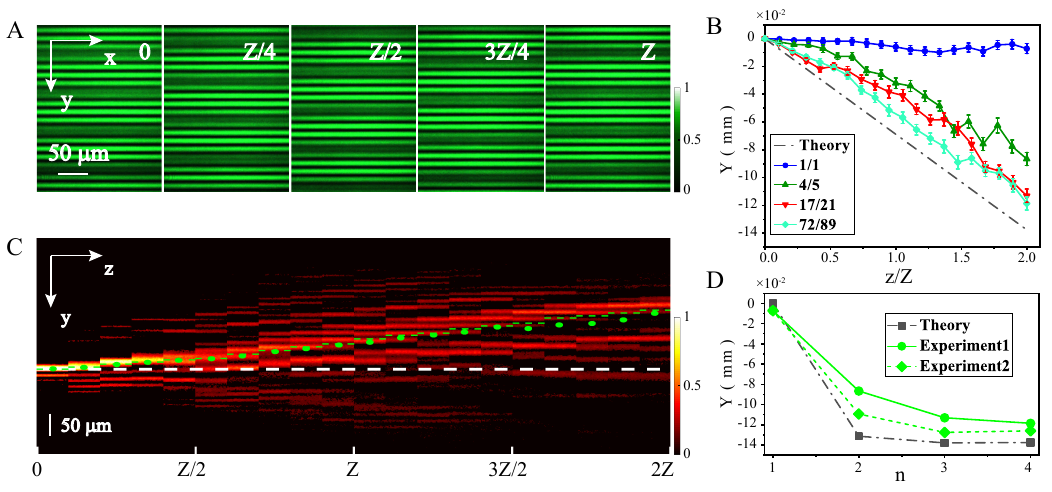}
\caption{Observation of light pumping in the lattice emulating a quasi-periodic structure with $\varphi=(\sqrt{5}+1)/4$. (A) The intensity distributions of the lattice obtained in BRA of the $4$th order ($p_4/q_4=72/89$) at specific distances $z=0,~Z/4,~Z/2,~3Z/4$ and $Z$. (B) Experimentally measured COM displacement versus distance $z$, represented by dots, and compared with the theoretically predicted curve shown as a dot-dashed line. (C) The intensity distribution of the probe beam on the $(y, z)$ plane, measured with steps of $2Z/19$ ($\approx1$ mm) in the propagation direction. The green dots indicate the positions of the COM at each interval of $2Z/19$, and the white dashed line indicates the position $y=0$, where the incident beam was launched. (D) The displacement of the COM at $z=2Z$, in the lattices corresponding to BRAs of the first four orders denoted by $n$. Curve “experiment 1” stands for the calculation of the COM from the whole diffraction field, while “experiment 2” stands for the calculation of the COM by discarding the contribution of the field below 20\% of the peak intensity. In all cases, $Z\approx 9.5~\text{mm}$, although it slightly varies with the order of rational approximations.}
\label{fig:3}
\end{figure*}

\end{document}